# Some considerations about snow crystallogenesis

**F.L. Falcon**

*Materials Science and Technology Institute, University of Havana, Cuba*

**Abstract**

We investigate about the possibility of knowing the thermal history of each snow crystal through the analysis of its individual habitus. Supposition, based on experimental observations, that prevailing growth mechanisms of basal and prismatic surfaces are helicoidal and 2D nucleation-spread, respectively, make possible to establish the relation temperature-habitus for all the different kinds of crystals, with the exception of plates in the interval -3º C < T < 0º C, where probably the surface melting plays an important role on the habitus development.

## 1. <u>INTRODUCTION</u>

Nakaya [1] has expressed that "the snow crystal is a hieroglyph from the heavens". This assertion promotes the question about the possibility of finding a biunivocal relation between the sequence of environmental conditions that have influenced the development of any snow crystal during its descent across different atmospheric strata and its individual habitus.

In this paper, we attempt to find the "codes" of the relation history-habitus through the analysis of some evident characteristics of this crystal growth process, in order to bring to light the prevailing phenomena capable of generate that infinite variety of capricious shapes.

## 1.1. Basal and prismatic surfaces growth mechanisms

The majority of snow crystals show on their basal surfaces a central growth terrace and frequently a hollow core with variable dimensions. These characteristics are conducive to assume that growth on these surfaces is preferably due to the development of helicoidal dislocations generated during the first stage of crystal nucleation. Actually, when snow crystal falls along an atmospheric layer, where the pressure of water vapor is lower than its saturation value with respect to ice on basal plane, this surface suffers an etching process, which is more intense on dislocation cores. Due to high density of these crystal defects along basal direction [2-5], the etching can become very intense, causing the formation of negative crystals and giving occasion for the appearance of interesting crystal shapes (i.e., hollow columns and hollow cups).

On the other hand, the strict symmetric growth of six prismatic surfaces on each snow crystal shows the negligible influence of dislocations on their growth. This fact and the tendency to dendritic growth on these surfaces induce to thinking about that their prevailing growth mechanism is the nucleation and further spread of 2D terraces.

## 2. **DISCUSSION**

### 2.1. Symmetric distribution of crystal habitus with temperature

In Fig.1, we denote the different snow crystals habitus and the atmospheric temperature range where each of them prevails. Furthermore, we denote the excess pressure on ice of water vapor for each temperature, which is the driving force for snow crystallization, according with Bergeron-Findeisen [6-7] model. At first glance we can observe in this figure the near symmetrical distribution of plates and columns (or needles) in the temperature range -30 $^0$C < T < -3 $^0$C, with respect to T~ -14 $^0$C as central temperature value.

This symmetrical distribution of habitus can be explained taking into account the different water vapor supersaturation dependences of helicoidal dislocation and 2-D growth mechanisms, which we have supposed prevail on basal and prismatic crystal surfaces, respectively.

It is known that helicoidal dislocation mechanism starts to work from lower super saturations than 2-D nucleation. On the other hand, a comparison of the density and relative strength of surface free bonds on basal ($0\ 0\ 0\ 1$) and prismatic ($0\ 1\ \bar{1}\ 0$) and ($1\ 0\ \bar{1}\ 0$) surfaces (Fig.2) shows that, after its initiation, the growth on prismatic surfaces must become more intense than on basal ones.

Analysis of Fig.1 shows that initiation of 2D growth on prismatic faces takes place above an excess vapor pressure over ice at water saturation ~ 0.20 hPa, corresponding to appearance of plate crystals.

According with former considerations and taking into account the symmetrical character of excess vapor pressure over ice at water saturation with temperature, we propose a relation between basal and prismatic surfaces normal growth rates and temperature, shown in Fig.3. Due to temperature influence on kinetic processes, the showed curves cannot be exactly symmetrical, because some decrease of growth rates at lower temperatures. This relation between basal and prismatic growth rates fully explains the different snow crystal habitus found at $T < -3\ ^0C$.

## 2.2 Hollow columns and needles

Another disturbing factor of symmetry in curves of Fig.3 is the existence of a greater number of hollow columns (or needles) at $-10\ ^0C < T < -3\ ^0C$ with respect to $T < -22\ ^0C$ temperature interval.

Frank [8] showed theoretically that axis of helicoidal dislocations with large Burgers vectors should take the form of hollow tubes, in order to minimize strains in the crystal structure. Such hollow cores, besides snow crystals, have been observed optically in alpha alumina whiskers [9] and using SEM in synthetic phlogopite [10]. As shown by Frank, the approximate equilibrium radius of the hollow axial core is related to the length of dislocation Burgers vector through the equation:

$$r = \frac{\mu b^2}{8\pi^2 \gamma} \qquad (1)$$

Where μ is the rigidity modulus of the material, b is the dislocation Burgers vector, and γ the surface energy of the material.

According to (1), at higher temperatures the equilibrium radius must increase due to surface energy decrease.

In addition, the density of structural defects in snow crystals that falls across warmer atmospheric layers must increase, because those crystals undergo more abrupt temperature variations in comparison with those ones that fall across more constant (cooler) temperature layers.

The two precedent considerations can explain the existence of a higher number of hollow columns (or needles) at -10 $^0$C < T < -3 $^0$C with respect to T < -22 $^0$C temperature interval.

## 2.3 Existence of plates in the range -3º C < T < 0º C.

The existence of plates in this temperature range cannot be explained by means of before exposed model. This non-agreement with proposed model probably can be explained by the appearance, slightly below the ice melting point, of a quasiliquid layer (QLL) at the ice-vapor interface [11-16]. Elbaum [17] has measured the QLL thickness as a function of temperature for the basal (0001)-vapour interface using optical reflectivity. Coincidently with the temperature corresponding to the change needles-plates in Fig.1, this author have obtained the minimum QLL thickness (1 nm) at T= -3$^0$C. The appearance of a QLL at the ice-vapour interface changes the growth crystal mechanism from vapour-solid to vapour-liquid-solid, whose analysis is out of the scope of this paper.

## 2.4 Abrupt variation in crystal habitus from needles to plates at T~ -3º C

This abrupt habitus change supports the assumption that after nucleation in upper troposphere (T< -40º C) snow crystals size will maintain very little during their descent until arriving to bottom layers of atmosphere, where water vapor pressure is much higher. In these atmospheric bottom layers the crystals rapidly reach their final size and habitus in correspondence with environmental temperature, shown in Fig.1.

According with before, the minute crystals nucleated in upper atmosphere must have habitus similar to those ones collected in South Pole [18] in winter season, where temperature remains almost the same along all their descent.

## 2.5 Prismatic surfaces growth modeling

Before to describe how the proposed growth mechanisms can explain the existence of complicated secondary habitus, we need make some considerations about the growth behavior of snow crystal prismatic surfaces.

With this aim, we utilized the procedure exposed in [19, 20] for modeling the molecular growth process on this crystal surface. In first place, we have found the hexagonal ice (ice Ih) growth cell (Fig.4). This figure shows in expanded form the occupied volume shape and the distribution and relative strength of intermolecular bonds of each water molecule in this crystalline phase. Fig.2 shows the arrangement of these growth cells in the ice Ih crystal and how they build basal ( 0 0 0 1 ) and two prismatic, ( 0 1 $\bar{1}$ 0 ) and ( 1 0 $\bar{1}$ 0 ), surfaces.

Fig.5 shows the result of modeling the nucleation of a 2D island on a prismatic surface. We have followed the same procedure as in [19], where starting with an attached to the surface initial cell, the others are sequentially added, with precedence for attachments where stronger surface bonds are satisfied.

As can be seen, the result of this modeling process shows that, after the attachment of a first water molecule, the attachment of a second one for the formation of a surface dimer is energetically favorable. Nevertheless, the probability of attachment of a third joint molecule to the incipient 2D island is the same as for the surface attachment of another isolated molecule disconnected from this island.

This fact means that there are not preferential places on prismatic surfaces for molecular attachment. This means that under excess of vapor pressure, the 2D nucleation on crystal surfaces must occurs preferably on crystal edges and corners, where mass and heat flows are more intense. If excess vapor pressure is not too high, the nucleated on edges 2D islands spread along all prismatic faces causing the formation of prismatic crystals (columns). If the nucleation on edges exceeds the spread rate, the capping tendency of certain crystals arises. For still higher vapor supersaturations, the nucleation takes place preferably on crystal corners (dendritic growth) causing the appearance of stellar crystals

## 2.6 Description of complicated secondary habitus (i.e., capped columns and capped cups)

Taking into account the turbulent airflows commonly existing in real atmosphere, the presented growth model also can explain different more complicated secondary habitus. Due to the existence of these flows, snow crystals can encounter along their path different lower and higher temperatures. These temperature (and therefore, excess water vapor pressure) variations cause the different combination of faceted or dendritic growth on each snow crystal that result in the infinite variety of habitus that we can find in any snow sample.

For example, Fig.6 shows a detailed explanation of the formation of capped columns and cups ("tsutsumi" crystals). In this figure are correlated the different section shapes found in these complicated crystal habitus along their basal axis and the possible sequences of environmental temperatures that can cause their variations.

## 3. CONCLUSIONS

The individual habitus of each snow crystal can be explained through a specific historical sequence for the relation between helicoidal dislocations and 2D nucleation growth mechanisms rates, which take place on basal and prismatic crystal surfaces, respectively. This specific growth rates relation sequence is a consequence of the individual thermal history of each crystal during its fall along troposphere, particularly across its bottom layers.

**Figure captions**

**Figure 1**: Relation between different snow crystal habitus, environmental temperature range where each of them prevails, and atmospheric excess pressure on ice of water vapor corresponding to each temperature.

**Figure 2**: Arrangement of water molecules (growth cells) in the ice Ih crystal, showing the distribution and relative strength of surface free bonds on basal ($0\ 0\ 0\ 1$) and prismatic ($0\ 1\ \bar{1}\ 0$) and ($1\ 0\ \bar{1}\ 0$) surfaces.

**Figure 3**: Proposed approximate relation between basal and prismatic surfaces normal growth rates in dependence of environmental temperature.

**Figure 4**: Hexagonal ice (ice Ih) growth cell in expanded form, showing its shape and the distribution and relative strength of intermolecular bonds of each water molecule in this crystalline phase.

**Figure 5**: Result of modeling the nucleation of a 2D island on a prismatic surface. After the energetically favorable attachment of a surface dimer (dark grey cells), the attachment of a third cell (light grey) is equally probable on any of six joint to 2D island positions (1-3 and symmetric positions) or on a disconnected from island position (4).

**Figure 6**: Correlation between the different cross-section shapes found in capped columns and capped cups ("tsutsumi" crystals) along their basal axis and the possible sequence of environmental temperatures that cause their variations.

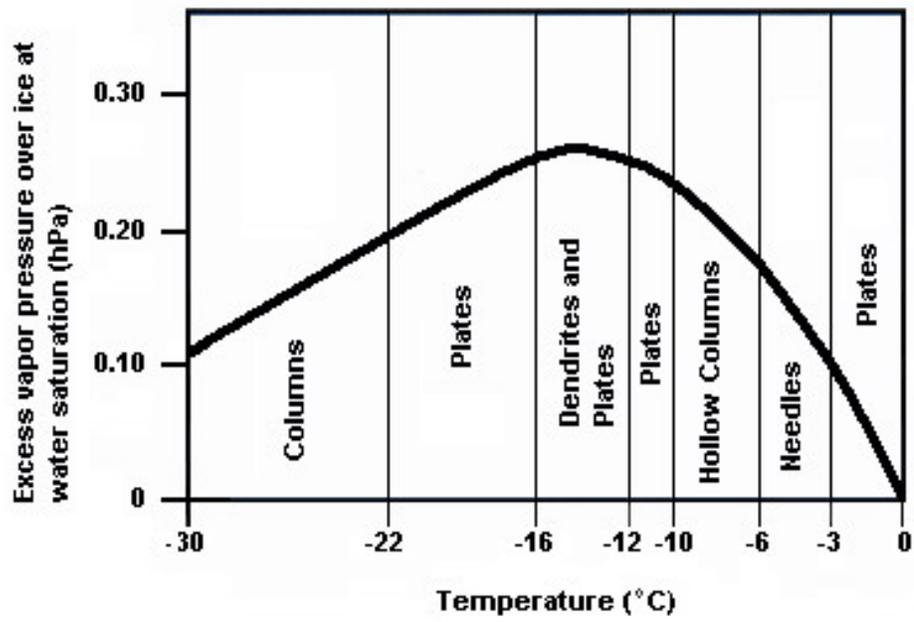

**Figure 1**

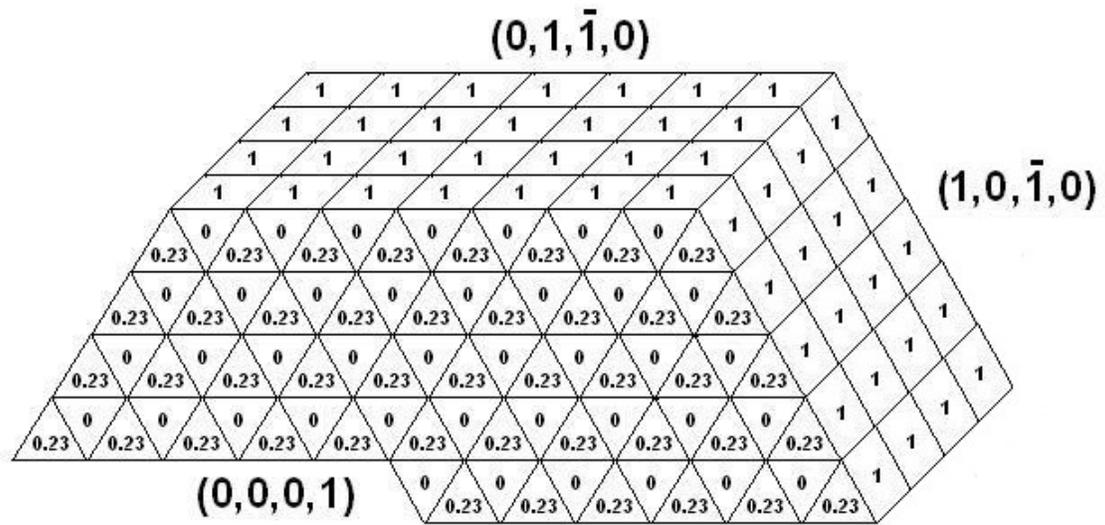

Figure 2

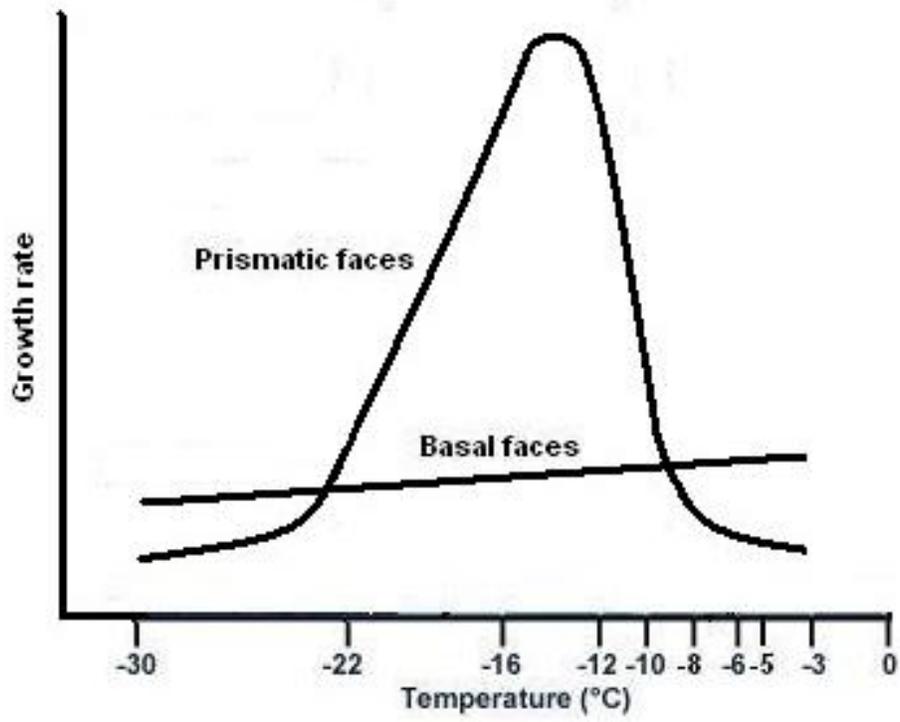

**Figure 3**

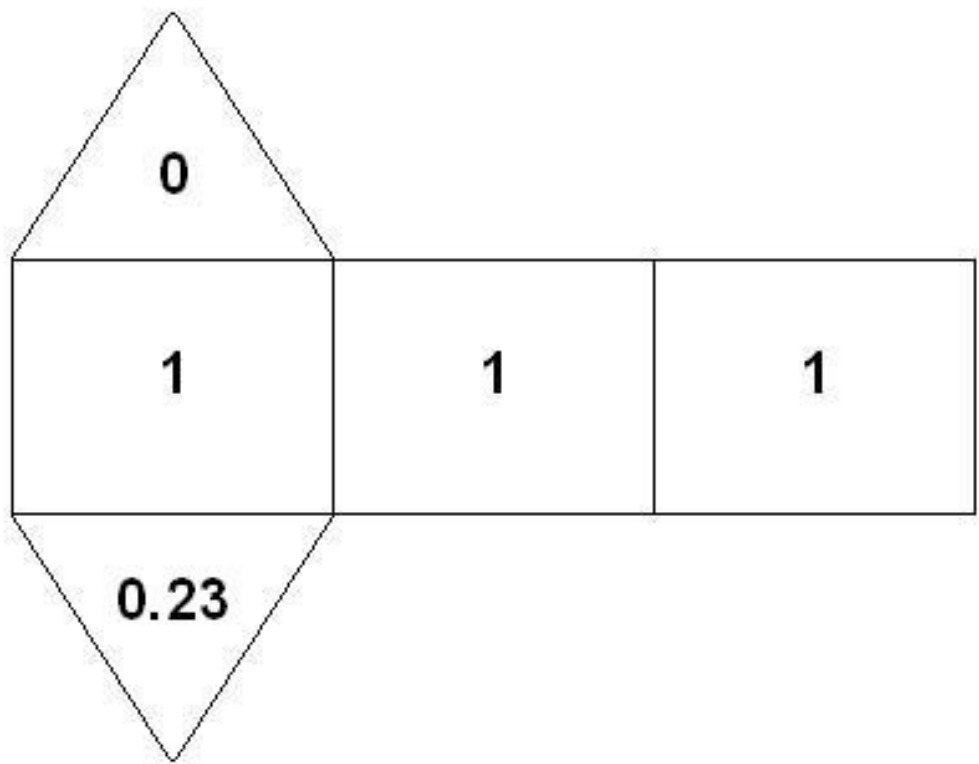

**Figure 4**

**Figure 5**

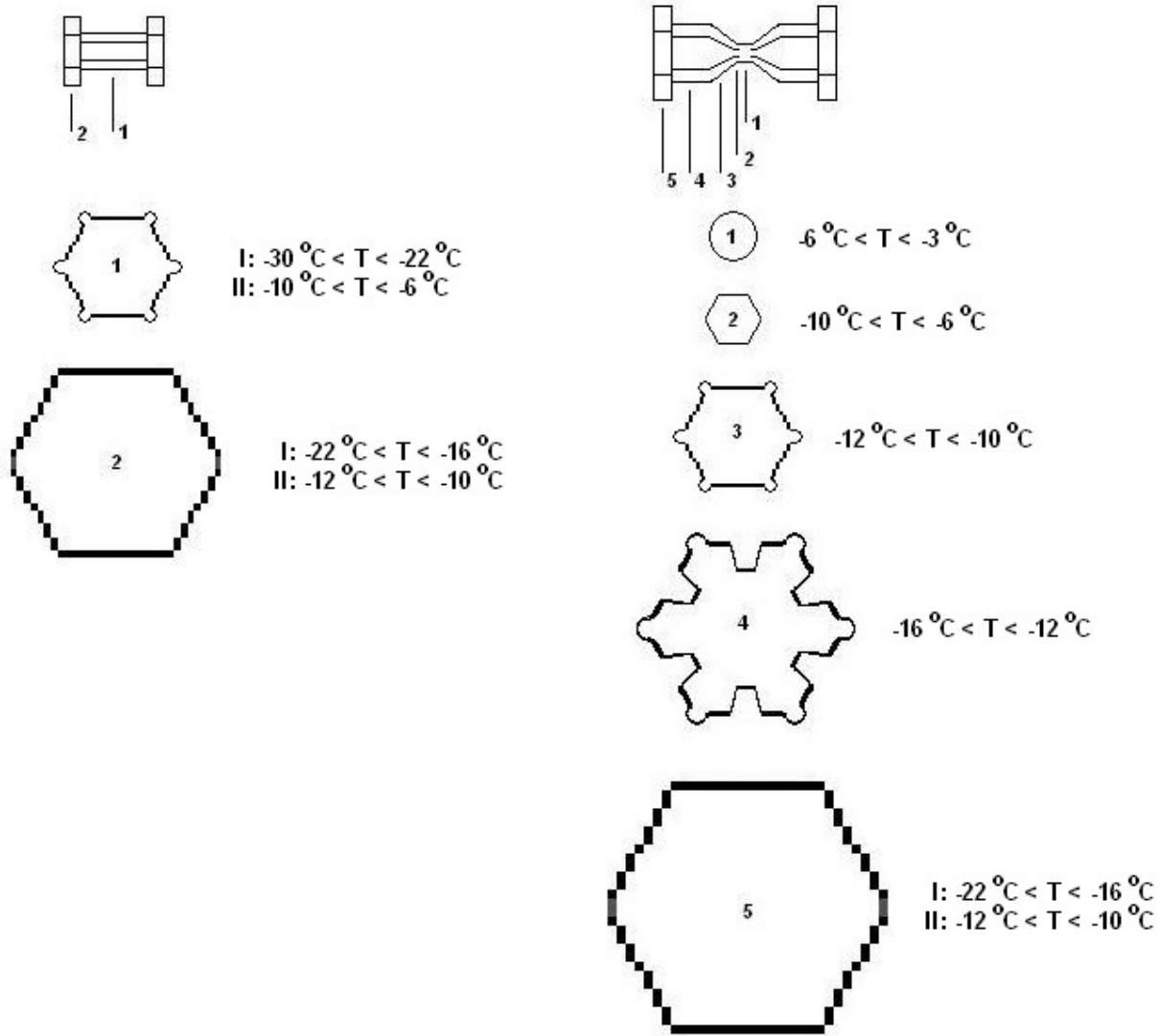

**Figure 6**